# Squeeze Excitation Embedded Attention UNet for Brain Tumor Segmentation

Gaurav Prasanna[1], John Rohit Ernest[1], Lalitha G[1], Sathiya Narayanan[2]

[1]School of Electronics Engineering, Vellore Institute of Technology, Chennai, India
[2]Centre for Advanced Data Science, Vellore Institute of Technology, Chennai, India

**ABSTRACT:**

Deep Learning based techniques have gained significance over the past few years in the field of medicine. They are used in various applications such as classifying medical images, segmentation and identification. The existing architectures such as UNet, Attention UNet and Attention Residual UNet are already currently existing methods for the same application of brain tumor segmentation, but none of them address the issue of how to extract the features in channel level. In this paper, we propose a new architecture called Squeeze Excitation Embedded Attention UNet (SEEA-UNet), this architecture has both Attention UNet and Squeeze Excitation Network for better results and predictions, this is used mainly because to get information at both Spatial and channel levels. The proposed model was compared with the existing architectures based on the comparison it was found out that for lesser number of epochs trained, the proposed model performed better. Binary focal loss and Jaccard Coefficient were used to monitor the model's performance.

1. **INTRODUCTION:**

Brain tumors are among the most dangerous types of tumors in the world. Brain tumors have proven difficult to treat, owing largely to the biological characteristics of these cancers, which frequently conspire to limit progress. The popular non-invasive technique called Magnetic Resonance Imaging (MRI) generates a large and diverse number of tissue contrasts in each imaging modality and has been broadly used by medical specialists to diagnose brain tumors. However, manual segmentation and analysis of structural MRI images of brain tumors is a complicated and time-consuming process that has been confined to professional neuroradiologists. As a result, brain tumor segmentation that is instantaneous and robust will have a considerable impact on brain tumor diagnosis and treatment. Such robust techniques include the existing CNN architectures like UNet, Attention UNet, and Attention ResUNet which are used for the purposes like biomedical image segmentation have done very well on benchmark datasets and have proven very vital even 3D Medical Imaging, but the major drawback all these three models face is none of them speak about or tackle when it comes to information at the channel level and that's the problem, we are trying to solve in this research work. In this paper we introduce a novel architecture called SEEA-UNet which stands for Squeeze Excitation Embedded Attention UNet, this model is a modification of existing Attention UNet to which Squeeze Excitation blocks are incorporated. The advantages this model proposes over the other models is that it has two main things to look out for firstly, incorporation of Attention UNet which proves very vital in feature extraction at focusing on the region of interest of the tumor is taken care by the Attention Gate and Gating Signal, and the second part is the embedding Squeeze Excitation blocks in encoding path of the modified UNet, which is very vital to analyze information at the channel level and this here makes our model Dual attention based mechanism. The dataset was taken from publicly available dataset from Kaggle. The Intersection over Union (IOU) score, also called as Jaccard index is one of the metrics that is used to evaluate the segmentation model and it is the most effective metric that is calculated from the area of overlap between the predicted segmentation and the ground truth divided by the area of union between the predicted segmentation and the ground truth. A greater IOU score results in accurate localization of the tumor. The rest of this paper is organized as follows; Section 2 discusses the existing methods that have been used for biomedical image segmentation. Section 3 discusses about the proposed architecture and its implementation, Section 4 discusses about the results and discussion, Section 5 discusses about the conclusion.

## 2. LITERATURE SURVEY:

In [1], a novel attention gate (AG) model was proposed for medical imaging that automatically learns to suppress irrelevant regions in an input image while highlighting salient features useful for a specific task. The Attention Gate was implemented in a standard U-Net architecture and was evaluated on the TCIA Pancreas CT-82 and multi-class abdominal CT-150 datasets. There was an improvement in prediction accuracy across different datasets and training sizes. In [2], a Recurrent Convolutional Neural Network as well as a Recurrent Residual Convolutional Neural Network (RRCNN) based on U-Net models was proposed. The model leverages U-Net, Residual Network, as well as RCNN. This model yielded a better accuracy as feature accumulation with recurrent residual convolutional layers ensured a better feature representation for segmentation tasks. The experimental results show superior performance on segmentation tasks compared to equivalent models including UNet and residual U-Net (Res-UNet). In [3], aggregated residual transformations were applied along with the soft attention mechanism on the computed tomography (CT) images to improve the capability in distinguishing a variety of symptoms of the COVID-19. This model enhanced feature extraction to a greater extent. There was a 10% improvement in multi class segmentation against U-Net and a set of baselines.

In [4], a novel Convolutional Neural Network (CNN), called USE-Net was proposed. It works by incorporating Squeeze - and - Excitation blocks into U-Net. The Squeeze Excitation blocks are added after every encoder or encoder-decoder block and were tested on MRI datasets. They found out that SE block's adaptive feature recalibration captures peculiar characteristics in a dataset to a quite a large extent. In [5], SERR-U-Net was proposed which leverages Squeeze and Excitation (SE), residual module and recurrent block. The convolution layers of encoder and decoder are modified on the basis of U-Net, and the recurrent block is used to increase the network depth. The residual module is utilized to alleviate the vanishing gradient problem. In order to derive more specific vascular features, the Squeeze Excitation structure was used to introduce attention mechanisms into the U-shaped network. This model improved the accuracy compared to learning based methods and its robustness in challenging cases is well demonstrated and validated. In [6], a unique low-parameter network based on 2D UNet architecture which employs two techniques, were proposed to prevent confusion of the model as well as computational complexity. An attention mechanism has been adopted after the concatenation of low-level and high-level features. The second technique is the introduction of multi-View Fusion that is adopted to benefit from 3D contextual information of input images instead of 2D model. In [7], a novel CNN architecture that exploits local features and global contextual features simultaneously and uses a fully connected final convolutional layer. This architecture also describes 2D training procedures to tackle imbalances in tumor labels. A cascade architecture is implemented which is designed such that the output of a fundamental CNN is fed as an additional source of information to the subsequent CNN. In [8], A 2D residual UNet with 104 convolutional layers, known as DR- UNet104 has been proposed for lesion segmentation of brain MRI images. The UNet architecture is added with residual blocks to the UNet encoder and dropout is added after each convolution block stack. The proposed architecture was examined on the Multimodal Brain Tumor Segmentation (BraTS)2020 challenge and was compared with DeepLabV3+ with a ResNet-V2-152 backbone. In [9], a novel segmentation framework called Segtran has been proposed which has unlimited effective receptive fields at higher feature resolutions. The experiments were performed on 2D as well as 3D medical image segmentation. The Segtran core is a novel Squeeze-and-Expansion transformer. A new positional encoding scheme for transformers has also been proposed in order to impose a continuity inductive bias for images. In [10], a novel deep network architecture called Skinny, a lightweight UNet has been employed with a wide spatial context with inception modules and dense blocks placed at every level of the network. Multiple Skinny networks have been trained over axial, coronal, and sagittal image planes and form an ensemble having such models.

## 3. SQUEEZE EXCITATION EMBEDDED ATTENTION UNet:

As mentioned in the Existing methods [18,19,21,23], most of the current methodologies are focused on spatial domain, that is all the operations for segmentation models are focused on two aspects that is the encoding and decoding path, where taking the example of UNet [11] Olaf Ronnerberger et al, says that the architecture consists of a contracting path to capture context and a symmetric path expanding path that enables precise location. This does an extremely good job in detecting the brain tumor and also localizing it. We Propose a novel idea where we focus on two aspects that is by modifying the UNet at two levels, first at the encoder path we have incorporate a Squeeze and Excitation Block, and secondly at the decoder path we have incorporated an existing modification done to UNet, that is the Attention UNet[1], by adding Gating Signal and Attention Gate(AG), the purpose of Attention gate and gating signal as said by Ozan Oktay et al is that while incorporating AG's for Image Analysis for having a better accuracy and segmentation they have come up with a way of integrating AG's with a standard Convolutional Neural Network(CNN) model, for converging to Region of Interest(ROI) the model need not have extra parameters to train also the requirement of Multiple models is also not needed, but in other manner speaking the AG's progressively suppress feature responses in irrelevant background regions without the requirement to crop a ROI between networks. And the gating signal is used in each skip connection along with the Attention Gate shown in fig 1, this is done so that it aggregates information from multiple imaging scales which increases the grid resolution of the query signal and achieves better performance.

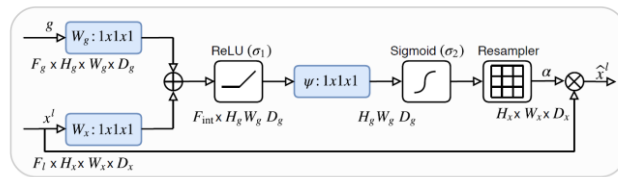

Fig 1: Schematic of the Attention Gate[1]

And apart from this the incorporation of Squeeze and Excitation Networks[12] (SE – Blocks) shown in fig 2, this was thought for because to further improve the existing performance of the network we have added SE Blocks in the encoding layers, as the names suggest there are two parts that is Squeeze operation and Excitation operation, the Squeeze part is the operation where the input image which consists of dimensions (H x W x C), this would be converted to (1 x 1 x C) and this done using Global Average Pooling (GAP) layer is used to compress the image and only give out the channel information to the Excitation layer. In Excitation layer it consists of a fully connected layer which is given along with a reduction factor r along to the ReLU activation and that is given to fully connected layer, and it finally goes through a sigmoid activation as the end result of Excitation layer. Now to feed the channel information obtained from the SE block the input image a scaling transformation is done and thus in the output we get information at the channel level, and this exactly is what we are trying to harness in the proposed architecture is to get best out of both the worlds, where UNet + Attention Gate can focus on Spatial part along with getting features necessary for the prediction without performing any hard image processing techniques like cropping. And Squeeze Excitation Blocks can give us information at the channel level, so that any important detail present in form of colour is not left out and fine attention is even given to that aspect.

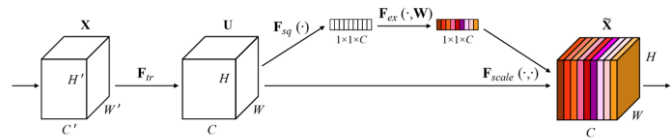

Fig 2: Depiction of Squeeze and Excitation Block[12]

In the proposed model that is Squeeze Excitation Embedded Attention UNet (SEEA-UNet), which is a modified version of UNet consisting of Squeeze Excitation blocks on Attention UNet, the total trainable parameters are 3.7 million parameters of which 15,618 are non-trainable parameters and rest 37,339,527 are trainable parameters. The modified Architecture is shown below in Fig 3.

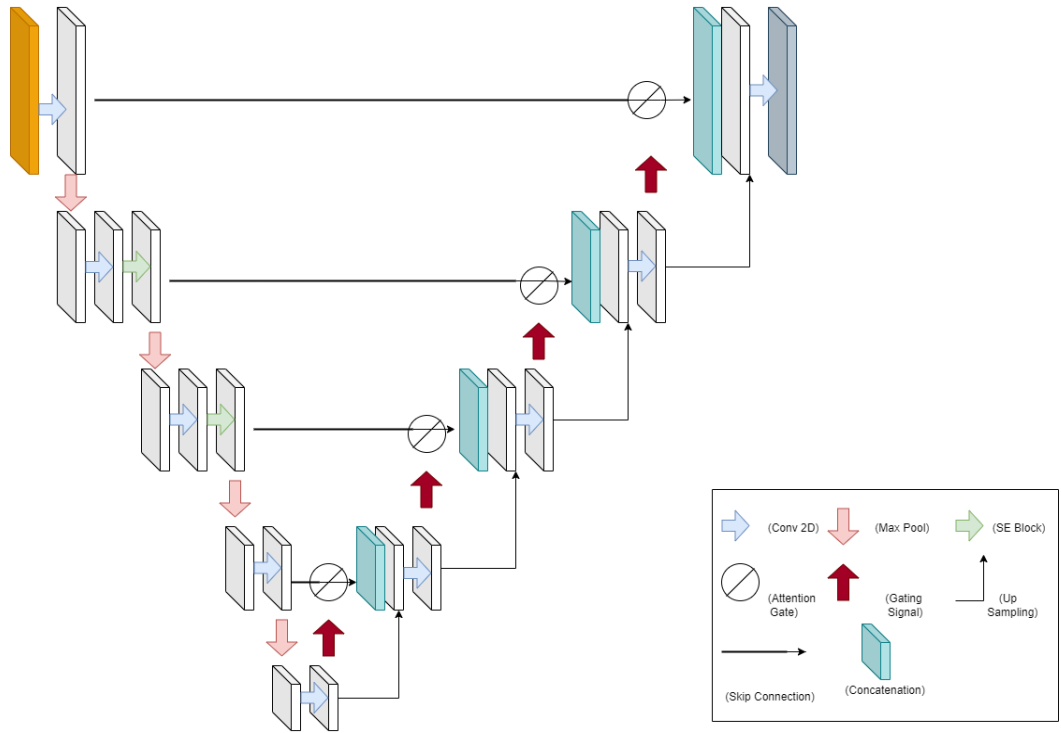

**Fig 3: Squeeze Excitation Embedded Attention UNet (SEEA-UNet)**

SEEA-UNet is extension of Modified Attention UNet, which promises to give better results with just increase of parameters of a very negligible number, the total trainable parameters of each of the models compared are given in the table 1.

**Table 1: Total and Trainable parameters for various models**

| CNN Model | Total Parameters | Trainable Parameters |
| --- | --- | --- |
| UNet | 31,401,349 | 31,389,571 |
| Attention UNet | 37,334,665 | 37,319,047 |
| Attention Residual UNet | 39,090,377 | 39,068,871 |
| Proposed SEEA-UNet* | 37,355,145 | 37,339,527 |

Features of SEEA-UNet which makes it a probable benchmark for not just brain tumor segmentation but also for any Biomedical Image Segmentation:

1) Though an extension of Attention UNet, the difference between total parameters in both have a very small difference, hence not leading to a wide gap in computational resources required to train the model, system used to train Attention UNet with same specifications can be used to train SEEA-UNet.
2) Able to gain information both at spatial domain and also at the channel levels, which gives significant chances for the model to learn better and able to extract features needed for tumor region along with Attention Gate with better ease.
3) The SEEA-UNet has been given the same hyperparameters as the other models to perform the comparison and study it how it performs along with other models.

## 4. RESULTS AND DISCUSSION:

The dataset used to train all the four models to perform the analysis was used from a publicly available dataset from Kaggle [13], the dataset consisted of 3929 images and masks of brain tumor, the Images were of RGB type images and corresponding masks were of Black and White type, due to availability of limited computational resources we handpicked a total of 700 images, from both the classes that is with tumor and without tumor. A sample image and mask are shown in the fig 4, the same data was used for other models to, All the four models were trained on Intel i5 12$^{th}$ generation processor with GPU by NVIDIA MX250 having a RAM of 8 GB.

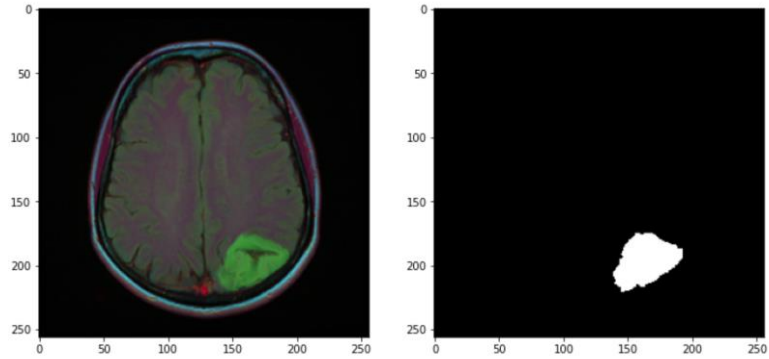

**Fig 4: Sample Image from the dataset and also its corresponding mask depicting the site of the tumor**

The hyperparameters used for the training are Adam optimizer with a learning rate of (0.01) and the loss function used here was Focal Loss (Binary Focal Loss) with gamma value of 2, the important metrics used to monitor the model's progress and also in general the performance of the model was checked with this along with the binary focal loss. The models were trained for 3 Epochs after which the model's performance started to degrade so we had to fix early stopping at 3 epochs, where the model yielded the best results, a detailed comparison between the Jaccard coefficients for various models is performed on Table 2, Jaccard Coefficient was used as the metric because IoU(Intersection over Union) is easy to monitor when it comes to segmentation results, overlaying the predicted mask with the ground truth mask would be a good understanding point on how well our model has performed and this is what the IoU does by finding out the common area between the two.

**Table 2: Jaccard Coefficient for all the four models at 3 Epochs and 5 Epochs**

| Name of the Model | Training Jaccard (3 Epochs) | Validation Jaccard (3 Epochs) | Training Jaccard (5 Epochs) | Validation Jaccard (5 Epochs) |
|---|---|---|---|---|
| UNet | 0.0301 | 0.0318 | 0.0450 | 0.0450 |
| Attention UNet | 0.0298 | 0.0256 | 0.0387 | 0.0553 |
| Attention Residual UNet | 0.0326 | 0.0232 | 0.0400 | 0.0485 |
| Proposed SEEA-UNET* | 0.0646 | 0.0721 | - | - |

**Table 3: Loss for all the four models trained with same hyperparameters at 3 Epochs**

| Model | Training Loss | Validation Loss |
|---|---|---|
| UNet | 0.0252 | 0.0203 |
| Attention UNet | 0.0252 | 0.0212 |
| Attention Residual UNet | 0.0241 | 0.0224 |
| Proposed SEEA-UNet* | 0.0097 | 0.0112 |

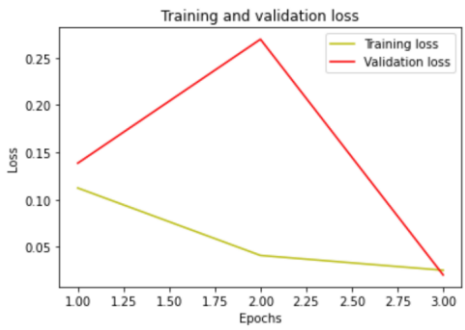
(a)
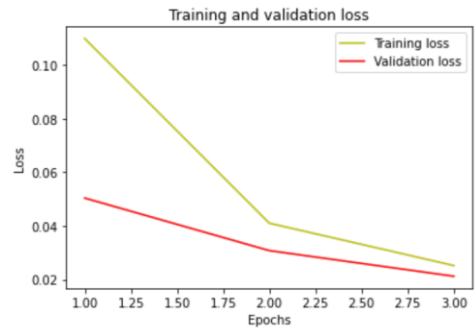
(b)
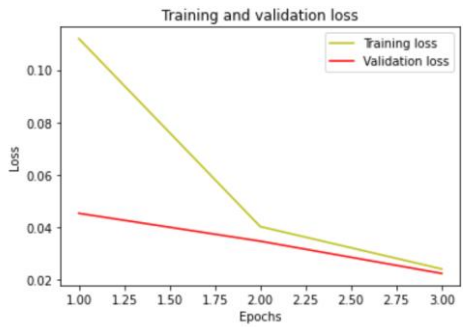
(c)
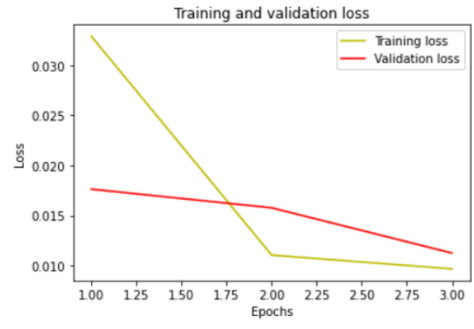
(d)

**Fig 5(a,b,c,d): (a,b,c) shows loss plot of UNet, Attention UNet and Attention Residual UNet Respectively and (d) depicts the loss proposed SEEA-UNet**

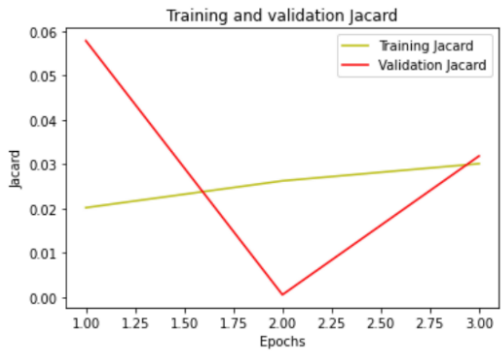
(a)
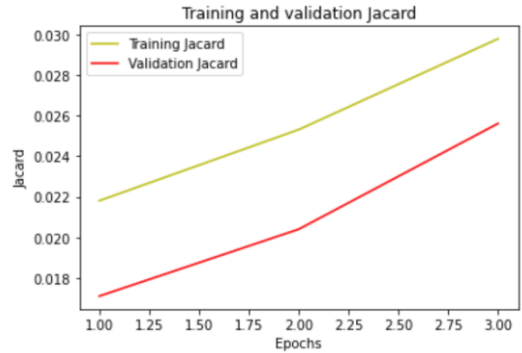
(b)
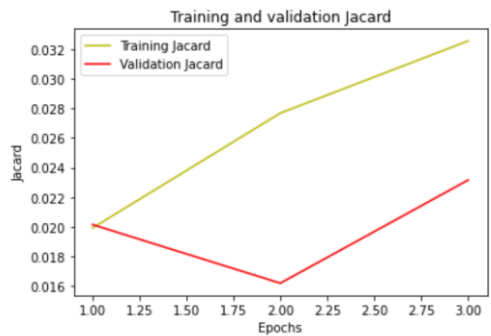
(c)
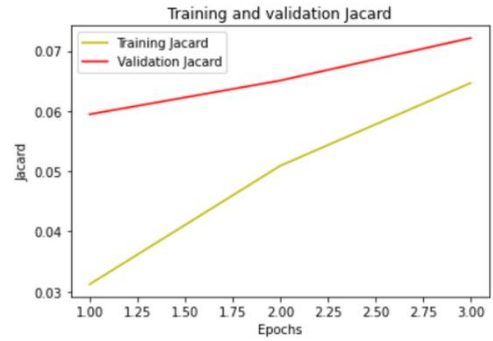
(d)

**Fig 6(a,b,c,d): (a,b,c) shows Jaccard plot of UNet, Attention UNet and Attention Residual UNet Respectively and (d) depicts the Jaccard of proposed SEEA-UNet**

From figures 5 and 6 also from tables 2 and 3 it is very clear that the binary focal loss is very less compared to the other three existing models, and even the Jaccard coefficient is higher than the other three models. So, with the current computational resources and dataset procured, the proposed SEEA-UNet can be benchmarked on this dataset.

## 5. CONCLUSION:

Overall, in this paper we try to prove that having a double attention-based scheme one that of Squeeze and Excitation Network based and other having soft attention mechanism by incorporating Attention gate and gating signal could very well serve as a benchmark for biomedical image segmentation. Not just solely focused on Brain Tumor Segmentation, addition to this could be incorporating more layers into the network and then test it in on different benchmark datasets and see how it performs on varied scales, and not just focusing on modified UNet. As further study this model can also be checked on 3D Medical Imaging Dataset and how well it is able to perform even in lesser number of epochs and varying the hyperparameters to see how it performs. on adding more layer this could very well lead to improvement in performance and aim at higher Jaccard coefficient and also bring the loss to a very negligible value.

ACKNOWLEDGEMENT:

Lastly, we would like to thank Dr Muthu Subhash Kavitha for the guidance and support along the way, and also to Dr Sreenivas Bhattiprolu from whom we had taken a lot code references from his tutorial and Github repository and also Research Scholar Premanand for his guidance and support while debugging the code.